\documentclass[%
aps,
pre,
 amsmath,amssymb,
 superscriptaddress,
preprint,%
longbibliography
]{revtex4-1}
\usepackage{graphicx}
\usepackage{dcolumn}
\usepackage{ulem}
\usepackage[utf8]{inputenc}
\usepackage[T1]{fontenc}
\usepackage{mathptmx}
\usepackage{textcomp} 
\usepackage{nicefrac}
\usepackage{xcolor}
\usepackage{gensymb}
\usepackage{makecell}
\usepackage{multibib}
\usepackage{multirow}

\usepackage{comment}

\usepackage{lineno}
\usepackage{csquotes}

\usepackage{ulem} 

\usepackage[framemethod=TikZ]{mdframed}
\usepackage{capt-of}

\definecolor{boxcolor1}{rgb}{0.87, 0.90, 0.84}
\definecolor{boxcolor2}{rgb}{0.93, 0.94, 0.91}

\mdfdefinestyle{MyFrame}{%
    frametitlebackgroundcolor = boxcolor1,
    hidealllines = true, 
    backgroundcolor=boxcolor2,
    roundcorner=1pt
        }

\usepackage[nopatch=footnote]{microtype}

\usepackage[breaklinks]{hyperref}
\newcommand{\tas}{\texorpdfstring{1\textit{T}-TaS\textsubscript{2}}{1\textit{T}-TaS2}}
\newcommand{\mos}{\texorpdfstring{2\textit{H}-MoS\textsubscript{2}}{2\textit{H}-MoS2}}
\newcommand{\feps}{\texorpdfstring{FePS\textsubscript{3}}{FePS3}}

\usepackage{siunitx}
\begin{document}

\title{Ultrafast order-selective electron imaging and spectroscopy}

\author{Jan Philipp Bange} \email{jbange@phys.uni-goettingen.de}%
\affiliation{I. Physikalisches Institut, Georg-August-Universit\"at G\"ottingen, Friedrich-Hund-Platz 1, 37077 G\"ottingen, Germany}

\author{Till Domröse} %
\email{till.domroese@mpinat.mpg.de}
\affiliation{Department of Ultrafast Dynamics, Max Planck Institute for Multidisciplinary Sciences, Am Fassberg 11, Göttingen, Germany}

\author{Claus Ropers} %
\affiliation{Department of Ultrafast Dynamics, Max Planck Institute for Multidisciplinary Sciences, Am Fassberg 11, Göttingen, Germany}
\affiliation{IV. Physikalisches Institut, Georg-August-Universit\"at G\"ottingen, Friedrich-Hund-Platz 1, 37077 G\"ottingen, Germany}

\author{Stefan Mathias}
\affiliation{I. Physikalisches Institut, Georg-August-Universit\"at G\"ottingen, Friedrich-Hund-Platz 1, 37077 G\"ottingen, Germany}
\affiliation{International Center for Advanced Studies of Energy Conversion (ICASEC), University of Göttingen, Göttingen, Germany}

\author{Marcel Reutzel}
\affiliation{I. Physikalisches Institut, Georg-August-Universit\"at G\"ottingen, Friedrich-Hund-Platz 1, 37077 G\"ottingen, Germany}
\affiliation{Fachbereich Physik, Philipps-Universität Marburg, 35032 Marburg, Germany}
\affiliation{mar.quest | Marburg Center for Quantum Materials and Sustainable Technologies, 35032 Marburg, Germany}

\begin{abstract}

Ultrafast pump–probe spectroscopies in energy and momentum have become indispensable for probing non-equilibrium dynamics in quantum materials, revealing pathways in ultrafast energy conversion and light-induced phase transitions. Different modalities uniquely access the coupled lattice, charge, and spin degrees of freedom that underpin the functionality of these materials. However, investigating technologically relevant nanoscale devices also requires high spatial resolution. Although electron microscopy provides nanometer-scale imaging, it frequently lacks the simultaneous ultrafast temporal resolution and spectroscopic specificity needed for such studies. In this perspective, we review recent advances in dark-field electron and photoelectron microscopy, focusing on two complementary techniques: femtosecond photoelectron momentum microscopy and ultrafast transmission electron microscopy. Their application enables comprehensive insights into the ultrafast dynamics of the electron, lattice, and spin subsystems. These order-selective electron imaging and spectroscopy techniques open broad scientific opportunities for a microscopic understanding of non-equilibrium phenomena in quantum materials, nanostructures and devices.

\end{abstract}

\maketitle

\section{Introduction}
The macroscopic properties of solids and nanostructures emerge from the intricate and dynamical interplay between their microscopic degrees of freedom: electrons, the lattice, and spins. Understanding this coupling underpins the development of new functionalities in next-generation technological applications, and ultrafast measurement methodology has substantially contributed to the characterization of the inherent dynamics on the timescale of femto- to picoseconds \cite{Petek12, delaTorre21Rev.Mod.Phys.}.  

The properties of crystalline solids are strongly defined by the periodic arrangement of the individual atoms into unit cells, i.e., the atomic lattice and the resulting structural and electronic ordering.
More subtle, topological properties and correlations that extend beyond unit-cell scales \cite{Keimer17NaturePhys} give rise to emergent phenomena spanning the subsystems: electronic correlations produce Wigner crystals, correlated insulators, and strong Coulomb-correlated excitations \cite{Mak22Nat.Nanotechnol., Regan22NatRevMater, Xu25npjQuantumMater.}, coupling to the lattice underlies charge-density waves \cite{Rossnagel11J.Phys.:Condens.Matter} and spins may form skyrmionic textures \cite{Finocchio16J.Phys.D:Appl.Phys.}.
By Fourier relation, such periodicities in real space typically manifest in characteristic fingerprints in reciprocal or momentum space.
Accordingly, a comprehensive description of an ordered system intrinsically relies on information resolved in momentum space $\mathbf{k}$, for example the quasi-particle energy dispersion 
$E(\mathbf{k})$.

In the past, spatially averaging, momentum-sensitive probes have greatly advanced the mechanistic understanding of condensed-matter systems and their non-equilibrium dynamics across a wide range of materials and model systems \cite{delaTorre21Rev.Mod.Phys.}. 
Specifically, time- and angle-resolved photoemission spectroscopy (trARPES) accesses the photoemission spectral function, which encodes information on the single-particle electronic band structure, many-body interactions, and electronic correlations \cite{Sobota21Rev.Mod.Phys., Boschini24Rev.Mod.Phys.}.
Momentum microscopy (MM) is a modern variant of trARPES that enables multidimensional measurements covering the full first Brillouin zone simultaneously \cite{Karni23Adv.Mater.}. This provides the complete momentum information efficiently in time-resolved studies \cite{Reutzel24Adv.Phys.X}. 
The unique advantage of MM is that it allows direct ultrafast band-structure mapping of previously inaccessible nanoscale samples \cite{Reutzel24Adv.Phys.X, Karni23Adv.Mater.}, e.g. exfoliated two-dimensional monolayers \cite{Aeschlimann25SurfaceScience}, opening up the strength of momentum-resolved studies to the fields of light-matter interaction, quasi-particle scattering dynamics and dissipation in hybrid interfaces and even optoelectronic devices \cite{delaTorre21Rev.Mod.Phys.}. 

In a complementary manner to photoemission, ultrafast electron diffraction (UED) elucidates the non-equilibrium temporal evolution of the lattice \cite{Filippetto22Rev.Mod.Phys.}. 
Time-dependent electron diffractograms trace the evolution of a material’s structural order, e.g., in the context of phase transitions \cite{Siwick03Science, Domrose23Nat.Mater.}, as well as coherent lattice oscillations \cite{Zong23Nature}, while inelastic scattering traces the temporal evolution of phonon populations \cite{Chase16Appl.Phys.Lett., Pan25ACSNano, Stern18Phys.Rev.B, Kurtz24Nat.Mater.}.

\begin{figure}
    \centering
    \includegraphics[width=1\linewidth]{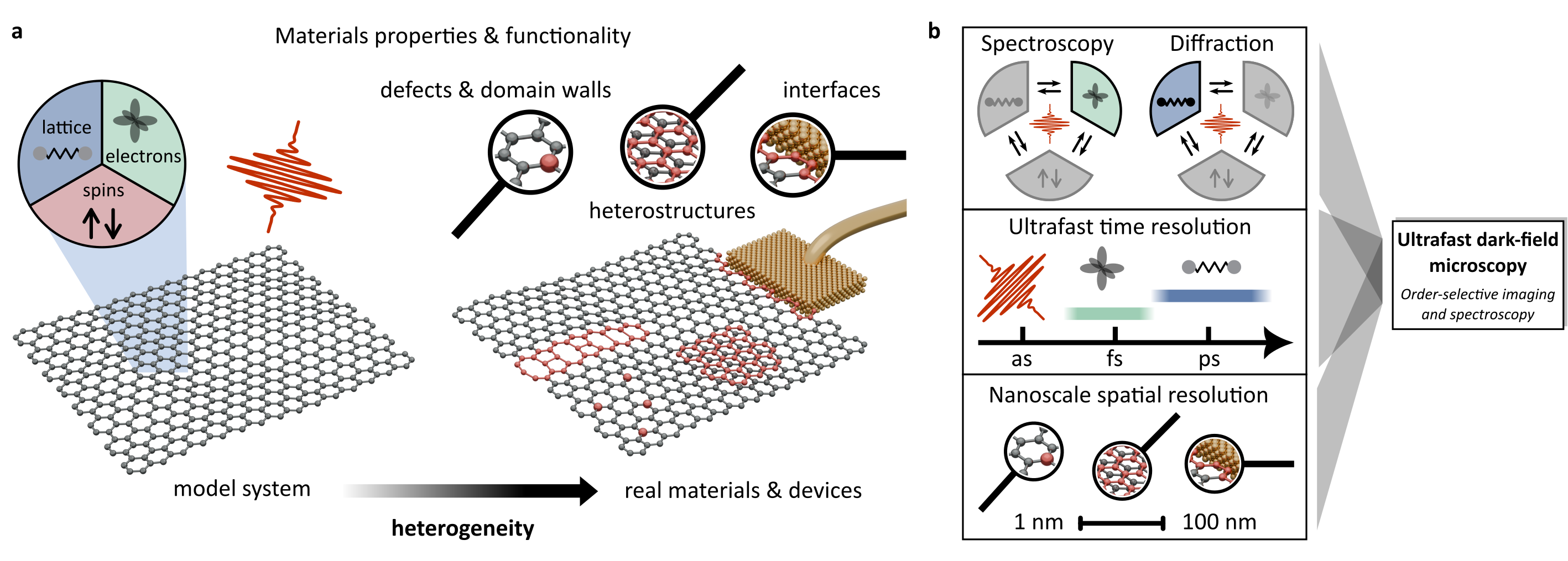}
    \caption{\textbf{Ultrafast order-selective imaging and spectroscopy}. \textbf{a} In real materials and functional devices beyond simple model systems, the ultrafast dynamics of lattice, electron, and spin subsystems must be analyzed in the context of heterogeneity. \textbf{b} A holistic understanding of ultrafast dynamical processes requires (1) high spatial resolution on the nanometer scale, (2) ultrafast time resolution, and (3) spectroscopic sensitivity in energy and momentum space.}
    \label{fig:overview}
\end{figure}

All these techniques, however, average over macroscopic sample areas. Real materials and devices naturally host defects and interfaces (Fig.\ref{fig:overview}a) in both pristine and heterogeneous structures \cite{Rhodes19Nat.Mater.}. While spatial heterogeneity can either contribute to or limit the functionality of technological applications \cite{Kennes21Nat.Phys., Rhodes19Nat.Mater., Lin162DMater., Haigh12NatureMater, Park16NanoLett., Raja19Nat.Nanotechnol.}, the impact on non-equilibrium electron-phonon-spin dynamics has remained largely inaccessible. In this context even less understood are device structures beyond model systems, i.e. operando devices, which are inherently heterogeneous and contain contacts and interfaces (Fig.\ref{fig:overview}a). In principle, the microscopy analogues of trARPES and UED, i.e., time-resolved photoemission electron microscopy (trPEEM) and ultrafast transmission electron microscopy (UTEM), offer the required nanoscale spatial sensitivity to femtosecond phenomena \cite{Dabrowski20Chem.Rev.}. Most implementations of these techniques, however, integrate over reciprocal space and are thereby insensitive to the rich information encoded in momentum-resolved observables. They are therefore not optimized to track a material’s phase (e.g., charge-density wave, Mott, superconducting states) and associated order parameters, band structure, momentum-specific many-body interactions, and their ultrafast dynamics. Consequently, the simultaneous \textit{spatio-temporal-spectral} imaging (Fig.\ref{fig:overview}b) of ultrafast dynamics remains a key objective, driving methodological advances that combine nanometer spatial and femtosecond temporal resolution with spectroscopic sensitivity to energy, momentum, or ideally both.

In this perspective, we describe recent advances in ultrafast MM and UTEM that realize momentum-selective nano-imaging of femtosecond dynamics. As exemplified by proof-of-principle experiments \cite{Danz21Science, vanderVeen13NatureChem, Cremons17Struct.Dyn., Schmitt25Nat.Photon., Paleschke25Phys.Rev.B, Maklar23UltrafastPhenom.NanophotonicsXXVIIb, Zhang19NanoLett.}, ultrafast dark-field microscopy schemes combine the momentum-selectivity of trARPES and UED with the nanometer spatial resolution of trPEEM and UTEM. This yields complementary information on electronic and structural dynamics in heterogeneous systems, respectively. Dark-field MM and UTEM promise new insights into ultrafast dynamics and energy-conversion processes in optoelectronic devices \cite{Liu16NatRevMater, Mao19J.Am.Chem.Soc., Wang18Chem.Soc.Rev.}, non-equilibrium structural and electronic phase transitions \cite{Basov17NatureMater, delaTorre21Rev.Mod.Phys.}, and the engineering of quantum materials' properties by light \cite{delaTorre21Rev.Mod.Phys., Disa21Nat.Phys.}.

The perspective is structured as follows: After a brief overview of experimental approaches combining ultrafast imaging with spectroscopic information (Sec.~II), we introduce the concept of ultrafast dark-field electron microscopy (Sec.~III). Secs.~IV and V present experimental realizations in MM and TEM, respectively. Finally, Sec.~VI offers a discussion of present limitations and an outlook on future advances in quantum materials' nanostructures and operando devices.

\section{Experimental pathways towards spatio-temporal-spectral resolution}

Nearly all existing approaches developed for achieving simultaneous temporal and spectral resolution at ultrafast timescales employ stroboscopic pump-probe schemes based on femtosecond laser pulses.
Depending on the observable, however, capabilities and limitations differ substantially. For instance, in all-optical microscopies, femtosecond light pulses are used to excite and probe the non-equilibrium dynamics by quantifying the sample's dielectric response \cite{Gross23J.Phys.Chem.C, Zhu19Annu.Rev.Phys.Chem., Maiuri20J.Am.Chem.Soc.},
albeit at diffraction-limited spatial resolution.
Consequently, high-energy photons generated from table-top high-harmonic generation beamlines or free-electron lasers offer x-ray absorption spectroscopy and imaging with high spatial resolution \cite{Vogelsang25EPL, Johnson23Nat.Phys., Zayko21NatCommun}. Complementarily, photons in the visible to terahertz (THz) spectral range can be confined to the nanoscale at atomically sharp metal tips \cite{Zhao25eLight}, culminating in the development of scanning near field optical microscopy (SNOM) \cite{Hillenbrand25NatRevMater, Cocker21Nat.Photon.} to ultrafast scanning tunneling microscopy (USTM) experiments \cite{Cocker16Nature, Muller24ProgressinSurfaceScience}. However, in both cases, access to the momentum-dependent spectral and structural correlations of the electron and lattice sub-systems is hardly possible \cite{Simon11J.Phys.D:Appl.Phys., Inbar23Nature}.

The combination of (photo-)electron spectroscopy and dark-field imaging, surveyed in the following, fills this gap. For broader introductions into the concepts of ultrafast imaging and spectroscopy, we refer to reviews of USTM \cite{Cocker21Nat.Photon., Muller24ProgressinSurfaceScience}, trPEEM \cite{Kosar24ProgressinSurfaceScience, Vogelsang25EPL, Lian23Int.J.Mech.Syst.Dyn., Dabrowski20Chem.Rev.}, trARPES \cite{Boschini24Rev.Mod.Phys., Echenique04Surf.Sci.Rep., Sobota21Rev.Mod.Phys., Bauer15Prog.Surf.Sci., Zhou18Rep.Prog.Phys., Aeschlimann25SurfaceScience}, trMM \cite{Karni23Adv.Mater.,Reutzel24Adv.Phys.X}, UED \cite{Filippetto22Rev.Mod.Phys.,Horstmann24ProgressinSurfaceScience}, and UTEM \cite{LaGrange25NatRevMethodsPrimers,Ji26Front.Phys.}.

\section{Dark-field electron spectroscopy in real space}

Dark-field imaging was developed in optical microscopy more than 100 years ago \cite{Gage20Trans.Am.Microsc.Soc.} and remains a standard technique in that field \cite{Gao21Anal.Chem., Liu14ColloidsandSurfacesB:Biointerfaces}.
The underlying principle of selective detection of scattered radiation applies naturally to electron microscopy \cite{Bauer94Rep.Prog.Phys., Beyer19Adv.Mater.Interfaces, Klein15Anal.Chem.}, regardless of whether the instrument is optimized for electron diffraction or ARPES.
Recent advances have extended dark-field imaging to ultrafast pump-probe experiments.
Depending on the type of experiment, elastic scattering or ARPES, an ultrashort light pulse photo excites a coherent electron probe beam from an electron gun \cite{Zewail10Science,Domer03Rev.Sci.Instrum.}, or photo emits electrons directly from the sample \cite{Boschini24Rev.Mod.Phys.}, respectively.

\begin{figure}
    \centering
    \includegraphics[width=1\linewidth]{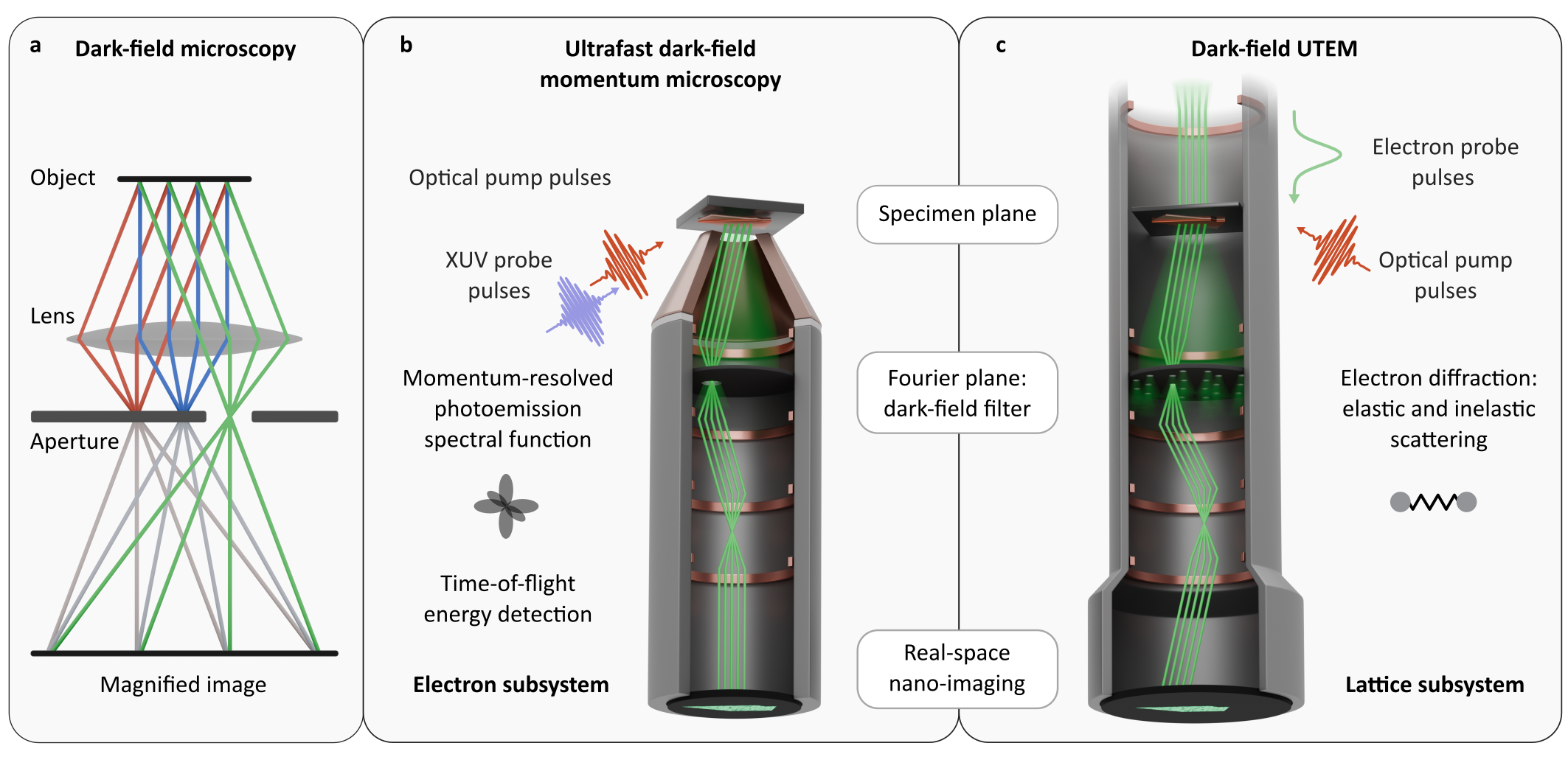}
    \caption{\textbf{Dark-field nano-imaging in ultrafast electron microscopy}
    \textbf{a}
    Concept of dark-field microscopy. 
    \textbf{b}
    Femtosecond momentum microscope: By placing an aperture in the back-focal plane of the microscope, one selectively filters the photoelectron spectral function in momentum space. The projection lens system subsequently transforms the image into real space with nanoscale spatial resolution.
    \textbf{c} Ultrafast transmission electron microscope (UTEM): A specific Bragg reflex in momentum space is selected for order-selective nano-imaging.}
    \label{fig:momi_utem}
\end{figure}

In its simplest form, dark-field filtering generates a selective contrast in the spatially resolved image on the detector. However, this approach's full potential is leveraged when combined with spectroscopic detection, enabling high-resolution spatial read-out of the spectral function. Capturing the spectral function in its entirety requires the dark-field aperture to be varied systematically in momentum space (Fig.~\ref{fig:momi_utem}b).

Analogously, in TEM and low-energy electron microscopy (LEEM), real-space structural order is mapped by positioning the aperture around different Bragg reflexes, as shown in Fig.~\ref{fig:momi_utem}c. Complementary four-dimensional scanning transmission electron microscopy (4D-STEM) achieves higher spatial resolution and detailed scattering information, yet relies on a real-space scanning approach \cite{Ophus19Microanal}.

\section{Femtosecond dark-field momentum microscopy}
\begin{figure}
    \centering
    \includegraphics[width=1\linewidth]{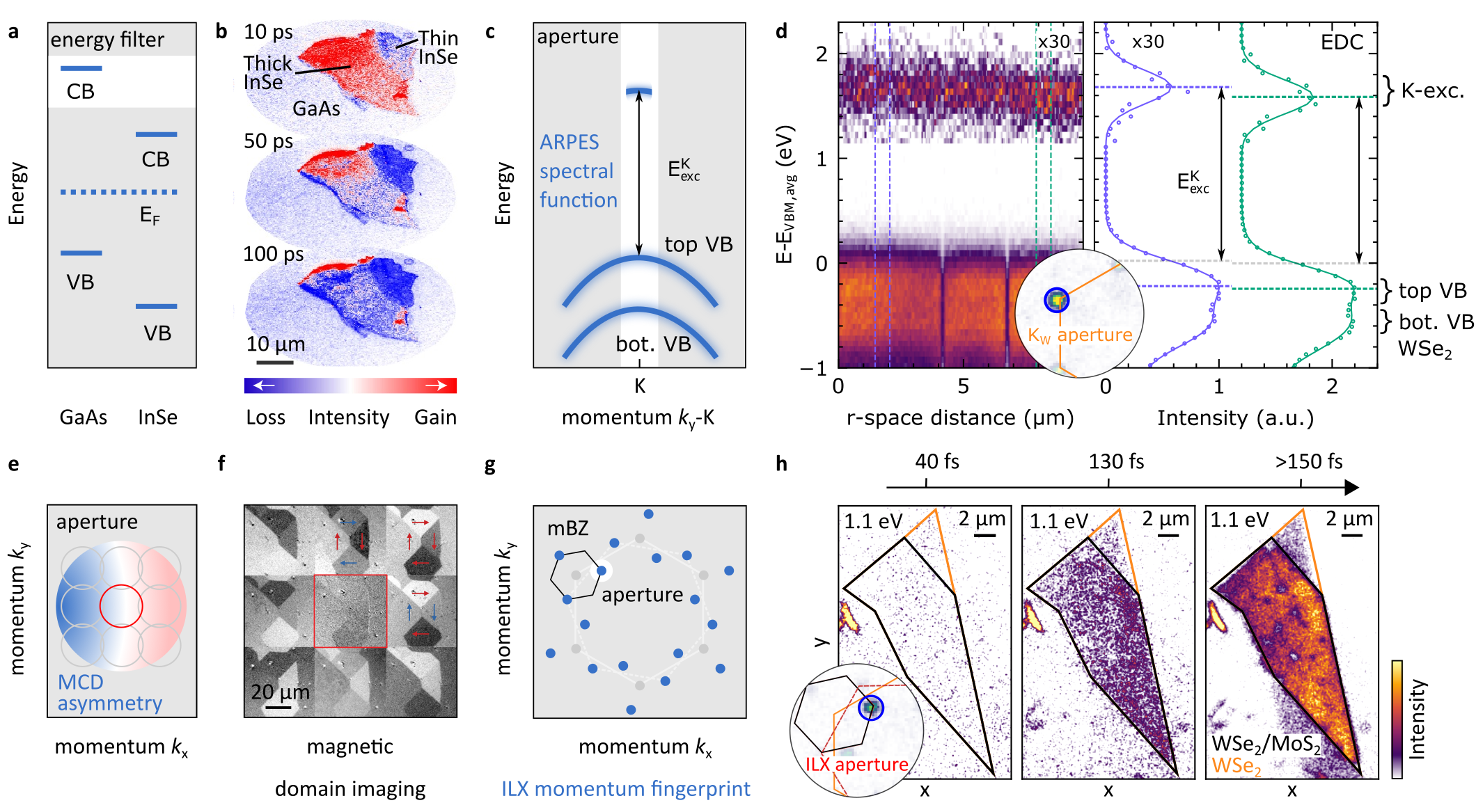}
    \caption{\textbf{Ultrafast dark-field photoelectron microscopy}
    \textbf{a} Photoelectron microscopy with an energy analyzer allows for spatio-spectral determination of valence band (VB) and conduction band (CB) edges in heterostructures.
    \textbf{b} Ultrafast motion of electrons across the interface of InSe and GaAs.
    \textbf{e}
    Dark-field aperture for momentum-resolved magnetic circular dichroism (MCD).
    \textbf{f}
    MCD contrast for imaging of magnetic domains. Individual panels correspond to respective aperture positions.
    \textbf{c}     
    Momentum-resolved spectral function of WSe$_2$ at the K-point, including top and bottom VB. The white cut-out indicates aperture position.
    \textbf{d} Nano-imaging of the non-equilibrium energy landscape of the excited heterobilayer WSe$_2$/MoS$_2$. Momentum filter is placed at the WSe$_2$ K-point (K$_{\mathrm{W}}$). Line profiles show the WSe$_2$ VB positions as well as the exciton signal (K-exc.).
    \textbf{g}
    Aperture (white circle) selects the momentum fingerprint of interlayer excitons (ILX) in the twisted heterobilayer, which resembles the moiré mini Brillouin zone (mBZ).
    \textbf{h}
    Ultrafast spatio-temporal formation of momentum-indirect ILX at $E-E_{\mathrm{ VBM}} = 1.1\,$eV on the heterobilayer area (black outlined).
        Parts of Fig. reproduced with permission from: \textbf{b} Ref.~\cite{Man17NatureNanotech} \& \textbf{d}, \textbf{h} Ref.~\cite{Schmitt25Nat.Photon.}, Springer Nature Limited, \textbf{e}, \textbf{f} Ref.~\cite{Paleschke25Phys.Rev.B}, APS.}
    \label{fig:spatio_spectral}
\end{figure}

trPEEM is a powerful tool to study ultrafast dynamical processes with exceptional surface sensitivity and spatial resolution below the optical diffraction limit \cite{Kosar24ProgressinSurfaceScience, Vogelsang25EPL}. 
Seminal experiments have focused on spatial-temporal characterization of surface plasmon polaritons (SPP) on nanostructured metal surfaces \cite{Kubo05NanoLett., Spektor17Science, Davis20Science, Dai20Nature, Hartelt21ACSNano}. 
More recently, trPEEM equipped with an energy analyzer captured the ultrafast transport of photo-excited charge carriers across an InSe/GaAs heterostructure (Fig.~\ref{fig:spatio_spectral}a,b) \cite{Man17NatureNanotech}.
Yet, access to the full information of the momentum-resolved spectral function, including the quasi-particle dispersion relations and many-body interactions \cite{Bostwick10Science}, is still limited.

trMM has emerged as a combination of PEEM and ARPES \cite{Kromker08Rev.Sci.Instrum., Medjanik17NatureMater}. Its ability to access the photoemission spectral function even on micrometer-sized flakes has opened up a rapidly expanding field of ultrafast photoelectron studies on monolayers of semiconductors and heterostructures \cite{Karni23Adv.Mater., Madeo20Science, Schmitt22Nature, Bange232DMater., Dong21Nat.Sci., Wallauer21NanoLett., Bange24Sci.Adv., Kunin23Phys.Rev.Lett., Bennecke24NatCommun, Bennecke25Nat.Phys.}.
Beyond this, in high-resolution real space mode, dark-field MM enables momentum filtering via a tunable aperture at the Fourier plane (Fig.~\ref{fig:spatio_spectral}c). 
Schmitt \textit{et al.} applied this method to image the ultrafast formation of interlayer excitons (ILX) in a WSe$_2$/MoS$_2$ heterostructure \cite{Schmitt25Nat.Photon.}.
In a first step, spatio-spectral dark-field imaging enabled real-space monitoring of the valence band alignment, a key feature of the photoemission spectral function. This directly helped to disentangle spatial inhomogeneities in the TMD's electronic band alignment, e.g., by placing the dark-field filter at the K-point of WSe$_2$ and following shifts in the spin-orbit split valence bands (Fig.~\ref{fig:spatio_spectral}d). 

Remarkably, this new approach can reveal band-shifts in samples with strain gradients or locally varying doping and defect concentrations. As demonstrated in the seminal work of Barrett \textit{et al.} \cite{Barrett12Rev.Sci.Instrum.}, it can also characterize systems with rotated domains, grain boundaries, and local reconstructions.
Furthermore, magnetic domains can be probed element-specifically by selecting the momentum-depended circular dichroism, extending the sensitivity to the spin subsystem (Fig.~\ref{fig:spatio_spectral}e,f) \cite{Paleschke25Phys.Rev.B}.

In a second step, ultrashort optical pump pulses access photoexcited states and the energy landscape far from thermal equilibrium. Accordingly, Fig.~\ref{fig:spatio_spectral}d shows nanoscale variations of exciton transition energies in twisted WSe$_2$/MoS$_2$. Upon optical excitation, Coulomb-correlated electron-hole pairs, i.e., excitons, form at the K-points of TMDs. The subsequent probe pulse photo-excites the exciton's electron, thereby dissociating the bound pair. These electrons are then detected one exciton energy above the valence band maximum and localized in momentum space at the K-points \cite{Bange24Sci.Adv., Bennecke24NatCommun, Weinelt04Phys.Rev.Lett.}, enabling their direct identification in the spatio-spectral information (Fig.~\ref{fig:spatio_spectral}d). Strikingly, this demonstrates that key features of the photoemission spectral function, e.g., energetic positioning of single particle bands, correlated quasiparticles, band-edge alignment, and broadening due to many-body renormalization \cite{Bostwick10Science, Ulstrup19NatCommun, Kastl19ACSNano, Wilson17Sci.Adv., Bao17NanoLett.} can be tracked in the non-equilibrium regime \cite{Duvel22NanoLett., Werner26Phys.Rev.Lett.}. 

Consequently, in the next step, dark-field nano-imaging enables tracing ultrafast dynamics in the non-equilibrium energy landscape of heterogeneous or intentionally structured samples. Fig.~\ref{fig:spatio_spectral}h depicts femtosecond snapshots of the ultrafast formation of interlayer excitons (ILX) in the WSe$_2$/MoS$_2$ heterostructure.
Surprisingly, pronounced spatial inhomogeneities in ILX formation dynamics are observed even in topographically flat regions, otherwise hidden for conventional all-optical techniques \cite{Mueller18npj2DMaterAppl}. 
Correlating the acquired exciton energy landscape (Fig.~\ref{fig:spatio_spectral}d) and the ILX formation time (Fig.~\ref{fig:spatio_spectral}h) sheds light on the underlying microscopic mechanisms of hybridization~\cite{Schmitt25Nat.Photon.}.
In a complementary manner, Maklar \textit{et al.} tracked the spatio-temporal evolution of a charge density wave (CDW) to metal transition in bulk TbTe$_3$. Buried defects were found to suppress the CDW order and the ultrafast response of collective states \cite{Maklar23UltrafastPhenom.NanophotonicsXXVIIb}.
Hence, the sensitivity to the momentum-resolved spectral function in combination with the ability of accessing both the non-equilibrium energy landscape and the ultrafast nanoscale dynamics is a dedicated strength of ultrafast dark-field momentum microscopy, providing unprecedented insights into quantum materials.

\clearpage
\section{Ultrafast dark field transmission electron microscopy}

\begin{figure}
    \centering
    \includegraphics[width=1\linewidth]{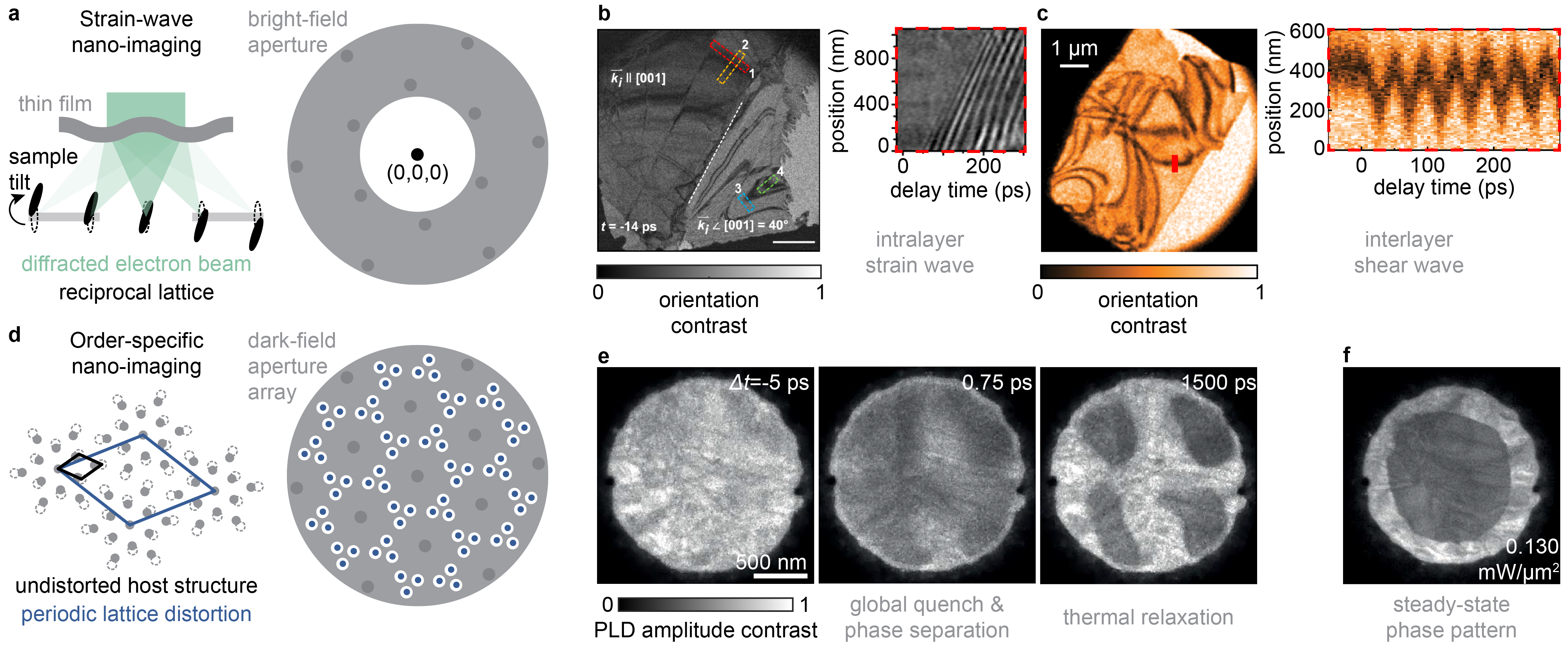}
    \caption{\textbf{Dark-field imaging in ultrafast transmission electron microscopy} \textbf{a} A single-aperture filter facilitates ultrafast strain mapping, visualizing coherent intralayer phonons (\textbf{b}) and interlayer shear modes (\textbf{c}) in laser-excited thin films. \textbf{d} Strong electron phonon coupling favors periodic lattice distortions (PLDs) into crystal superperiodicities (blue unit cell). Filtering for the associated satellite diffraction spots (blue) by a customized aperture array translates the PLD amplitude into nanoscale image contrast variations \textbf{(e)}. The ultrafast PLD phase transition strongly differs from the steady-state phase pattern emerging from continuous-wave laser excitation \textbf{(f)}. Parts of Fig. reproduced with permission from: \textbf{b} Ref.~\cite{Zhang19NanoLett.}, American Chemical Society, \textbf{c} Ref.~\cite{Zong23Nature}, Springer Nature Limited, \textbf{e}, \textbf{f} Ref.~\cite{Danz21Science}, AAAS.}
    \label{fig:utem}
\end{figure}

As outlined above, ultrafast electron diffraction (UED) elucidates the lattice component of laser-induced dynamics, involving the excitation of (coherent) phonons and the emergence or suppression of structural order \cite{Filippetto22Rev.Mod.Phys.}. Ultrafast transmission electron microscopy (UTEM) extends these observations to nanometer spatial resolution \cite{Zewail10Science, Domer03Rev.Sci.Instrum.}, leveraging the availability of imaging, diffraction, and spectroscopy within the same instrument. In its most common realization, femtosecond-to-picosecond electron pulses are generated via photoemission, allowing for stroboscopic laser-pump/electron-probe experiments \cite{LaGrange25NatRevMethodsPrimers,Ji26Front.Phys.}. Extending the capabilities of continuous-beam TEM, nanoscale imaging by Fourier optics is already widely employed in investigations of ultrafast structural dynamics. Figures \ref{fig:utem}a,b illustrate such an experiment, visualizing the emission of optically-induced coherent phonons in a \mos{} thin film \cite{Zhang19NanoLett.}. In electron diffractograms, the change in relative orientation between the electron beam and the thin film resulting from these lattice oscillations is recognizable by time-periodic modulations of spot intensities, including the effectively unscattered beam. A single bright-field aperture selecting this \enquote{direct} beam translates its intensity modulations into real-space imaging contrast, thus resolving the generation and propagation of phonons on the nanoscale.

Different modes can be distinguished by the temporal evolution of the image contrast. Intralayer phonons, launched at sample defects or edges, cross the field-of-view at the speed-of-sound (Fig.~\ref{fig:utem}b) \cite{Cremons17Struct.Dyn.,Nakamura20NanoLett.,Joshi25ACSNano}, whereas interlayer modes lead to a stationary intensity oscillation whose frequency is determined by the film thickness \cite{Joshi25ACSNano} (see Fig.~\ref{fig:utem}c for a shear wave excited in \feps{} \cite{Zong23Nature}). Repositioning the aperture in reciprocal space to allow for filtering based on a diffracted beam augments the sensitivity of phonons linked specifically to the selected crystallographic direction \cite{Ji22Nanoscale, Tong23Nat.Nanotechnol.}.

Besides sample orientation, diffracted intensities depend on the atomic positions within the material's unit cell, enabling bright- or dark-field imaging of structural phase transformations \cite{vanderVeen13NatureChem,Kim23Sci.Adv.,Pofelski26NatCommun}. Accessing the order parameter of such a phase switch directly, however, requires going beyond a single-aperture scheme. Often, the change of lattice symmetry is tied to specific wave vectors distributed across a larger $k$-space region. The formation of structural superperiodicities such as charge-density waves, for example, can be recognized by the suppression of characteristic diffraction spots while a set of new periodicities emerges \cite{Eichberger10Nature}. The associated periodic lattice distortions (PLD) typically involve sub-\r{A}ngstrom atomic displacements superimposed with the undistorted host lattice. Investigating these processes by DF microscopy then requires combining optimized signal transmission with high momentum selectivity, as imaging subtle structural order benefits from transmitting as many symmetry-equivalent diffraction spots as possible while minimizing the transmission of diffuse scattering, and avoiding $k$-space overlap with competing periodicities.

As a first realization of such a customized dark-field experiment, Figs.~\ref{fig:utem}d,e show the nano-imaging of a charge-density wave phase transformation in \tas{} \cite{Danz21Science}. At room-temperature, electron diffractograms of the material's nearly-commensurate (NC) CDW phase feature characteristic superstructure satellite spots around the bright reflexes of the undistorted host structure (blue colors in Fig.~\ref{fig:utem}d). The PLD spot intensity is a direct measure of the PLD amplitude, the structural order parameter of the CDW transition. Heating transforms the crystal into an incommensurate (IC) CDW, wherein the PLD wave vector slightly rotates by about \SI{12}{\degree}. This structural change translates into the extinction of the NC CDW spots and the emergence of IC reflections closely nearby.

For nanoimaging, the customized DF-filter features an array of \num{72} apertures with small diameters that block the high-temperature IC spots appearing after the transition. Ultrafast imaging then yields grayscale contrast that scales quadratically with the local NC PLD amplitude (Fig.~\ref{fig:utem}e). The combination of \num{5}\textendash\SI{}{\nm} spatial and sub\textendash\num{500}\textendash\SI{}{\fs} temporal resolution elucidates spatio-temporal dynamics of a heterogeneous phase transformation that reaches beyond the spatially-averaged information obtained by UED measurements. Dynamics in the NC and IC phase domains can be evaluated individually in terms of the local dark and bright image contrast. A clear phase separation sets in already after \SI{2}{\ps}, followed by the slower, nanosecond relaxation back to the initial NC phase governed by thermal diffusion that involves a temporary small growth of the IC domains. Additionally, the uniform initial contrast darkening along the comparably thick \tas{} specimen in the IC regions suggests rapid hot-carrier transport along the film depth.

\clearpage
\section{Advances in ultrafast electron microscopy} 
\subsection{Challenges and Limitations}

The main experimental challenge of ultrafast dark-field imaging is an intrinsic trade-off between resolution and obtainable contrast. Coulomb repulsion broadens and elongates the ultrashort pulses that contain more than one electron. Single-electron probe pulses maximize the temporal resolution and both transverse and longitudinal coherence \cite{Aidelsburger10PNAS}, but at the expense of signal. The coherent electron current achievable in UTEM measurements is typically limited by the employed electron source. Laser-driven tip emitters offer particularly high beam coherence \cite{Feist17Ultramicroscopy, Zhu20Ultramicroscopy, Schroder25Ultramicroscopy, Kuttruff24Sci.Adv., Houdellier18Ultramicroscopy, Olshin20Struct.Dyn.}, thus enabling high momentum resolution in nanoscale probe volumes. In MM, the sample itself acts as the electron source, and Coulomb repulsion particularly deteriorates the energy resolution by shifting and distorting the spectrum \cite{Schonhense18NewJ.Phys.}. This effect is especially present in time-resolved studies where also pump pulses emit bunches of slow secondary electrons.

Ultrafast DF microscopy, however, critically depends on high temporal sensitivity, the distinction of closely related periodicities, and their selective filtering both in momentum and energy space, while only a fraction of the available signal carries information on the non-equilibrium evolution of electronic and structural ordering. The spatiotemporal mapping of high-intensity diffracted signals such as strain waves can be conducted at few-\SI{}{\kHz} repetition rates. In contrast, following low-intensity order parameters directly requires enhanced duty cycles, i.e., hundreds of kilohertz to megahertz MM setups \cite{Puppin19Rev.Sci.Instrum., Keunecke20Rev.Sci.Instrum., Allison25APLPhotonics}. Thermal management schemes in UTEM measurements have been shown to dissipate the deposited energy required for driving structural dynamics within nanoseconds \cite{Domrose25Appl.Phys.Lett.}. Further experimental flexibility is provided by rotatable DF filter holders that ease the usage of aperture arrays \cite{Domrose24NanoLett.}. Accounting for the expected long image acquisition times (\SI{1}{\hour} per image at a repetition rate of \SI{420}{\kHz} for the dataset displayed in Fig.~\ref{fig:utem}d), an advantage of direct imaging is the simultaneous sampling of the entire field-of-view that enables an efficient drift correction. Such large-area movies of the dynamics may be complemented by ultrafast four-dimensional scanning transmission electron microscopy, wherein a focused electron nanobeam is rastered across the sample \cite{Nakamura22FaradayDiscuss.,Niedermayr26ACSNano}. Recording a diffractogram at every scan position in specific regions of interest characterized by, e.g., interfaces, domain walls, or the stacking of different materials promises to combine the phonon-sensitive imaging displayed in Figs.~\ref{fig:utem}a-c with the phase sensitivity shown in Figs.~\ref{fig:utem}d,e.

\subsection{Future Outlook and Applications: Nanostructured devices}
In order to unravel the interactions between the electron, phonon, and spin subsystems at the nanoscale (Fig.~\ref{fig:overview}), order-specific and momentum-resolved ultrafast imaging is now available, as described above. However, imaging the spin subsystem selectively with electron-based methods remains a challenge. In that regard, spin-selective multidimensional scattering filters are developed for spin-polarization-resolved band mapping \cite{Schonhense15JournalofElectronSpectroscopyandRelatedPhenomena, Schonhense17Ultramicroscopy}. Magnetic linear dichroism MM experiments have already visualized sub-micrometer-sized antiferromagnetic domains \cite{Fedchenko22J.Phys.:Condens.Matter}.

In strongly correlated materials, the subsystems are intertwined \cite{Hellmann12NatCommun}, often leading to nanoscale heterogeneity, in which small rearrangements in one of them cause significant modifications of the others. The resulting macroscopic properties are complex, including competing ground state phases and hidden states \cite{Stojchevska14Science}. A comprehensive understanding requires not only resolving the heterogeneity but also probing both the electron and the lattice system simultaneously \cite{Gerber17Science}. Consequently, a combination of dark-field trMM and UTEM promises to disentangle the contribution of collective excitation dynamics \cite{Xu25npjQuantumMater.} with dynamical band renormalizations. While trMM is highly surface sensitive, making it an ideal tool for studying single atomic layers, UTEM probes the structural dynamics of entire thin films. In contrast, static (dark-field) LEEM provides structural information and surface sensitivity. Extending dark-field LEEM in the ultrafast time domain will therefore enable these complementary time-resolved photoemission and structural dynamics experiments.

\begin{figure}
    \centering
    \includegraphics[width=1\linewidth]{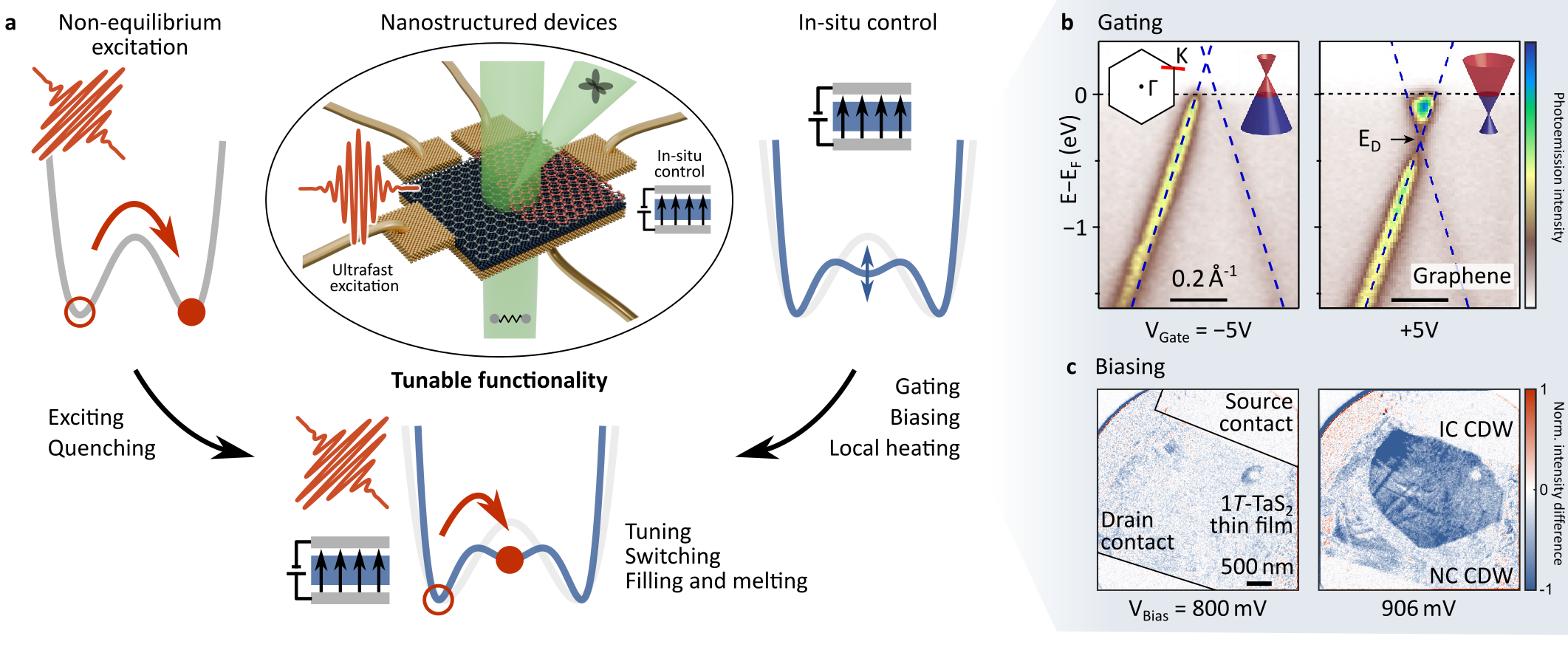}
    \caption{\textbf{Tunable emergent functionality in nanostructured devices}
    \textbf{a} In-situ control of nanostructured quantum materials and excitation of non-equilibrium dynamics leads to tunable emergent functionality.
    \textbf{b} Photoemission spectroscopy visualizes in-situ electric field gating of monolayer graphene.
    \textbf{c} Current-induced switching from nearly commensurate (NC) to incommensurate (IC) charge density waves (CDW) in \tas{}.
    Parts of Fig. reproduced with permission from \textbf{b} \cite{Nguyen19Nature}, Springer Nature Limited, \textbf{c} \cite{Domrose24NanoLett.}, American Chemical Society.
    }
    \label{fig:outlook}
\end{figure}

Beyond investigations of material properties under out-of-equilibrium conditions, ultrafast order-selective imaging is particularly beneficial for the characterization and development of nanoscale devices that feature external tuning of material properties (Fig. \ref{fig:outlook}). Typically, such structures feature engineered spatial heterogeneity in the form of, e.g., electrical contacts or the stacking of different materials. Then, functional control promises to drastically extend the range of transient or metastable states available after driving the system out-of-equilibrium. These schemes may include modifying the system’s ground state by displacement fields and temperature gradients, manipulating the population of electronic states in complex free-energy landscapes by electric gating, or harnessing functionality only accessible by combining optical excitation with electrical biasing.
For example, the ultrafast melting and recovery of correlated electronic phases at integer fillings in moiré sites will elucidate competitions between electron correlations and electron-phonon interactions, as well as the interplay with structurally reconstructed domains \cite{Arsenault24Phys.Rev.Lett., Duncan25Nature}. Even further, correlated states at fractional fillings might only be transiently accessible \cite{Wang25Nature}.
On the structural side, ultrafast nanoimaging will elucidate spatially heterogeneous phase transitions driven not only electrically \cite{Svetin17SciRep, Yoshida15Sci.Adv., Domrose24NanoLett.} or by light \cite{Eichberger10Nature,Danz21Science}, but also a combination thereof, thus resolving, e.g., optical switching in the presence of CDW sliding induced by below-threshold biasing \cite{Brown23Appl.Phys.Rev.}. 

In summary, we have presented ultrafast dark-field nano-imaging in trMM and UTEM as a powerful technique for investigating order-specific and momentum-resolved fundamental dynamical processes. The advantages of these techniques are given by the ultrafast spatio-temporal-spectral imaging capabilities, which promise to deliver unprecedented insights into dynamics in heterogeneous quantum materials in the future.

\clearpage
\section*{Acknowledgments}
A special thanks goes to David Schmitt, Wiebke Bennecke, Matthijs Jansen, and the entire Göttingen photoemission team: Mattis Langendorf, Junde Liu, Hashima Marukara, Marco Merboldt, Daniel Steil, Bent van Wingerden, and Paul Werner. Furthermore, we thank the Göttingen UTEM team for continued support.
We acknowledge support with artistic renders by Lukas Kroll.
This work was funded by the Deutsche Forschungsgemeinschaft (DFG, German Research Foundation) - Research Grant (Project No. 567214325), 217133147/SFB 1073 (projects A05, B07, and B10), the Priority Program SPP 2244 “2DMP” (Project No. 535247173), and via resources from the Gottfried Wilhelm Leibniz Prize (RO 3936/4-1). 
\section*{Competing interests}
The authors declare no competing interests.
\clearpage
\bibliographystyle{Science_jp_v2}
\bibliography{bibtexfile}

@article{Aeschlimann25SurfaceScience,
  title = {Time-Resolved Photoelectron Spectroscopy at Surfaces},
  author = {Aeschlimann, Martin and Bange, Jan Philipp and Bauer, Michael and Bovensiepen, Uwe and Elmers, Hans-Joachim and Fauster, Thomas and Gierster, Lukas and H{\"o}fer, Ulrich and Huber, Rupert and Li, Andi and Li, Xintong and Mathias, Stefan and Morgenstern, Karina and Petek, Hrvoje and Reutzel, Marcel and Rossnagel, Kai and Sch{\"o}nhense, Gerd and Scholz, Markus and Stadtm{\"u}ller, Benjamin and St{\"a}hler, Julia and Tan, Shijing and Wang, Bing and Wang, Zehua and Weinelt, Martin},
  year = {2025},
  month = mar,
  journal = {Surface Science},
  volume = {753},
  pages = {122631},
  issn = {0039-6028},
  doi = {10.1016/j.susc.2024.122631},
  urldate = {2024-11-27}
}

@article{Bauer15Prog.Surf.Sci.,
  title = {Hot Electron Lifetimes in Metals Probed by Time-Resolved Two-Photon Photoemission},
  author = {Bauer, M. and Marienfeld, A. and Aeschlimann, M.},
  year = {2015},
  month = aug,
  journal = {Progress in Surface Science},
  volume = {90},
  number = {3},
  pages = {319--376},
  issn = {0079-6816},
  doi = {10.1016/j.progsurf.2015.05.001},
  urldate = {2024-07-10}
}

@article{Boschini24Rev.Mod.Phys.,
  title = {Time-Resolved {{ARPES}} Studies of Quantum Materials},
  author = {Boschini, Fabio and Zonno, Marta and Damascelli, Andrea},
  year = {2024},
  month = feb,
  journal = {Reviews of Modern Physics},
  volume = {96},
  number = {1},
  pages = {015003},
  publisher = {American Physical Society},
  doi = {10.1103/RevModPhys.96.015003},
  urldate = {2024-07-03}
}

@article{Cocker21Nat.Photon.,
  title = {Nanoscale Terahertz Scanning Probe Microscopy},
  author = {Cocker, T. L. and Jelic, V. and Hillenbrand, R. and Hegmann, F. A.},
  year = {2021},
  month = aug,
  journal = {Nature Photonics},
  volume = {15},
  number = {8},
  pages = {558--569},
  publisher = {Nature Publishing Group},
  issn = {1749-4893},
  doi = {10.1038/s41566-021-00835-6},
  urldate = {2025-03-04},
  copyright = {2021 Springer Nature Limited},
  langid = {english}
}

@article{Danz21Science,
  title = {Ultrafast Nanoimaging of the Order Parameter in a Structural Phase Transition},
  author = {Danz, Thomas and Domr{\"o}se, Till and Ropers, Claus},
  year = {2021},
  month = jan,
  journal = {Science},
  volume = {371},
  number = {6527},
  pages = {371--374},
  publisher = {American Association for the Advancement of Science},
  doi = {10.1126/science.abd2774},
  urldate = {2023-05-11}
}

@article{Echenique04Surf.Sci.Rep.,
  title = {Decay of Electronic Excitations at Metal Surfaces},
  author = {Echenique, P. M. and Berndt, R. and Chulkov, E. V. and Fauster, {\relax Th}. and Goldmann, A. and H{\"o}fer, U.},
  year = {2004},
  month = may,
  journal = {Surface Science Reports},
  volume = {52},
  number = {7},
  pages = {219--317},
  issn = {0167-5729},
  doi = {10.1016/j.surfrep.2004.02.002},
  urldate = {2024-07-10}
}

@article{Gage20Trans.Am.Microsc.Soc.,
  title = {Modern {{Dark-Field Microscopy}} and the {{History}} of {{Its Development}}},
  author = {Gage, Simon Henry},
  year = {1920},
  journal = {Transactions of the American Microscopical Society},
  volume = {39},
  number = {2},
  eprint = {3221838},
  eprinttype = {jstor},
  pages = {95--141},
  publisher = {[American Microscopical Society, Wiley]},
  issn = {0003-0023},
  doi = {10.2307/3221838},
  urldate = {2025-04-14}
}

@article{Gao21Anal.Chem.,
  title = {Dark-{{Field Microscopy}}: {{Recent Advances}} in {{Accurate Analysis}} and {{Emerging Applications}}},
  shorttitle = {Dark-{{Field Microscopy}}},
  author = {Gao, Peng Fei and Lei, Gang and Huang, Cheng Zhi},
  year = {2021},
  month = mar,
  journal = {Analytical Chemistry},
  volume = {93},
  number = {11},
  pages = {4707--4726},
  publisher = {American Chemical Society},
  issn = {0003-2700},
  doi = {10.1021/acs.analchem.0c04390},
  urldate = {2025-04-14}
}

@article{Horstmann24ProgressinSurfaceScience,
  title = {Structural Dynamics in Atomic Indium Wires on Silicon: {{From}} Ultrafast Probing to Coherent Vibrational Control},
  shorttitle = {Structural Dynamics in Atomic Indium Wires on Silicon},
  author = {Horstmann, Jan Gerrit and B{\"o}ckmann, Hannes and Kurtz, Felix and Storeck, Gero and Ropers, Claus},
  year = {2024},
  month = jun,
  journal = {Progress in Surface Science},
  volume = {99},
  number = {2},
  pages = {100743},
  issn = {0079-6816},
  doi = {10.1016/j.progsurf.2024.100743},
  urldate = {2025-04-16}
}

@article{Kosar24ProgressinSurfaceScience,
  title = {Time-Resolved Photoemission Electron Microscopy of Semiconductor Interfaces},
  author = {Kosar, Sofiia and Dani, Keshav M.},
  year = {2024},
  month = sep,
  journal = {Progress in Surface Science},
  volume = {99},
  number = {3},
  pages = {100745},
  issn = {0079-6816},
  doi = {10.1016/j.progsurf.2024.100745},
  urldate = {2025-03-04}
}

@article{Kubo05NanoLett.,
  title = {Femtosecond {{Imaging}} of {{Surface Plasmon Dynamics}} in a {{Nanostructured Silver Film}}},
  author = {Kubo, Atsushi and Onda, Ken and Petek, Hrvoje and Sun, Zhijun and Jung, Yun S. and Kim, Hong Koo},
  year = {2005},
  month = jun,
  journal = {Nano Letters},
  volume = {5},
  number = {6},
  pages = {1123--1127},
  publisher = {American Chemical Society},
  issn = {1530-6984},
  doi = {10.1021/nl0506655},
  urldate = {2025-03-03}
}

@article{Lian23Int.J.Mech.Syst.Dyn.,
  title = {Probing Electron and Lattice Dynamics by Ultrafast Electron Microscopy: {{Principles}} and Applications},
  shorttitle = {Probing Electron and Lattice Dynamics by Ultrafast Electron Microscopy},
  author = {Lian, Yiling and Sun, Jingya and Jiang, Lan},
  year = {2023},
  journal = {International Journal of Mechanical System Dynamics},
  volume = {3},
  number = {3},
  pages = {192--212},
  issn = {2767-1402},
  doi = {10.1002/msd2.12081},
  urldate = {2025-02-28},
  copyright = {{\copyright} 2023 The Authors. International Journal of Mechanical System Dynamics published by John Wiley \& Sons Australia, Ltd on behalf of Nanjing University of Science and Technology.},
  langid = {english}
}

@inproceedings{Maklar23UltrafastPhenom.NanophotonicsXXVIIb,
  title = {Ultrafast Spatiotemporal Dynamics of a Charge-Density Wave Using Femtosecond Dark-Field Momentum Microscopy},
  booktitle = {Ultrafast {{Phenomena}} and {{Nanophotonics XXVII}}},
  author = {Maklar, J. and Walmsley, P. and Fisher, I. R. and Rettig, L.},
  year = {2023},
  month = mar,
  eprint = {2304.00839},
  primaryclass = {cond-mat},
  pages = {15},
  doi = {10.1117/12.2649985},
  urldate = {2025-03-04},
  archiveprefix = {arXiv}
}

@article{Man17NatureNanotech,
  title = {Imaging the Motion of Electrons across Semiconductor Heterojunctions},
  author = {Man, Michael K. L. and Margiolakis, Athanasios and {Deckoff-Jones}, Skylar and Harada, Takaaki and Wong, E Laine and Krishna, M. Bala Murali and Mad{\'e}o, Julien and Winchester, Andrew and Lei, Sidong and Vajtai, Robert and Ajayan, Pulickel M. and Dani, Keshav M.},
  year = {2017},
  month = jan,
  journal = {Nature Nanotechnology},
  volume = {12},
  number = {1},
  pages = {36--40},
  issn = {1748-3387, 1748-3395},
  doi = {10.1038/nnano.2016.183},
  urldate = {2022-02-03},
  langid = {english}
}

@article{Muller24ProgressinSurfaceScience,
  title = {Imaging Surfaces at the Space--Time Limit: {{New}} Perspectives of Time-Resolved Scanning Tunneling Microscopy for Ultrafast Surface Science},
  shorttitle = {Imaging Surfaces at the Space--Time Limit},
  author = {M{\"u}ller, Melanie},
  year = {2024},
  month = feb,
  journal = {Progress in Surface Science},
  volume = {99},
  number = {1},
  pages = {100727},
  issn = {0079-6816},
  doi = {10.1016/j.progsurf.2023.100727},
  urldate = {2024-11-13}
}

@article{Schmitt25Nat.Photon.,
  title = {Ultrafast Nano-Imaging of Dark Excitons},
  author = {Schmitt, David and Bange, Jan Philipp and Bennecke, Wiebke and Meneghini, Giuseppe and AlMutairi, AbdulAziz and Merboldt, Marco and P{\"o}hls, Jonas and Watanabe, Kenji and Taniguchi, Takashi and Steil, Sabine and Steil, Daniel and Weitz, R. Thomas and Hofmann, Stephan and Brem, Samuel and Jansen, G. S. Matthijs and Malic, Ermin and Mathias, Stefan and Reutzel, Marcel},
  year = {2025},
  month = jan,
  journal = {Nature Photonics},
  volume = {19},
  number = {2},
  pages = {187--194},
  issn = {1749-4893},
  doi = {10.1038/s41566-024-01568-y},
  urldate = {2025-01-03},
  copyright = {2025 The Author(s), under exclusive licence to Springer Nature Limited},
  langid = {english}
}

@article{Sobota21Rev.Mod.Phys.,
  title = {Angle-Resolved Photoemission Studies of Quantum Materials},
  author = {Sobota, Jonathan A. and He, Yu and Shen, Zhi-Xun},
  year = {2021},
  month = may,
  journal = {Reviews of Modern Physics},
  volume = {93},
  number = {2},
  pages = {025006},
  publisher = {American Physical Society},
  doi = {10.1103/RevModPhys.93.025006},
  urldate = {2024-07-10}
}

@article{Vogelsang25EPL,
  title = {Attosecond Microscopy ---{{Advances}} and Outlook},
  author = {Vogelsang, J. and Mikkelsen, A. and Ropers, C. and Gaida, J. H. and Garg, M. and Kern, K. and Miao, J. and Schultze, M. and Ossiander, M.},
  year = {2025},
  month = feb,
  journal = {Europhysics Letters},
  volume = {149},
  number = {3},
  pages = {36001},
  publisher = {{EDP Sciences, IOP Publishing and Societ{\`a} Italiana di Fisica}},
  issn = {0295-5075},
  doi = {10.1209/0295-5075/adaf51},
  urldate = {2025-04-08},
  langid = {english}
}

@article{Zhou18Rep.Prog.Phys.,
  title = {New Developments in Laser-Based Photoemission Spectroscopy and Its Scientific Applications: A Key Issues Review},
  shorttitle = {New Developments in Laser-Based Photoemission Spectroscopy and Its Scientific Applications},
  author = {Zhou, Xingjiang and He, Shaolong and Liu, Guodong and Zhao, Lin and Yu, Li and Zhang, Wentao},
  year = {2018},
  month = apr,
  journal = {Reports on Progress in Physics},
  volume = {81},
  number = {6},
  pages = {062101},
  publisher = {IOP Publishing},
  issn = {0034-4885},
  doi = {10.1088/1361-6633/aab0cc},
  urldate = {2025-04-14},
  langid = {english}
}

@article{Kastl19ACSNano,
  title = {Effects of {{Defects}} on {{Band Structure}} and {{Excitons}} in {{WS}}{\textsubscript{2}} {{Revealed}} by {{Nanoscale Photoemission Spectroscopy}}},
  author = {Kastl, Christoph and Koch, Roland J. and Chen, Christopher T. and Eichhorn, Johanna and Ulstrup, S{\o}ren and Bostwick, Aaron and Jozwiak, Chris and Kuykendall, Tevye R. and Borys, Nicholas J. and Toma, Francesca M. and Aloni, Shaul and {Weber-Bargioni}, Alexander and Rotenberg, Eli and Schwartzberg, Adam M.},
  year = 2019,
  month = feb,
  journal = {ACS Nano},
  volume = {13},
  number = {2},
  pages = {1284--1291},
  publisher = {American Chemical Society},
  issn = {1936-0851},
  doi = {10.1021/acsnano.8b06574},
  urldate = {2025-04-30}
}

@article{Wilson17Sci.Adv.,
  title = {Determination of Band Offsets, Hybridization, and Exciton Binding in {{2D}} Semiconductor Heterostructures},
  author = {Wilson, Neil R. and Nguyen, Paul V. and Seyler, Kyle and Rivera, Pasqual and Marsden, Alexander J. and Laker, Zachary P.L. and Constantinescu, Gabriel C. and Kandyba, Viktor and Barinov, Alexei and Hine, Nicholas D.M. and Xu, Xiaodong and Cobden, David H.},
  year = {2017},
  month = feb,
  journal = {Science Advances},
  volume = {3},
  number = {2},
  pages = {e1601832},
  issn = {2375-2548},
  doi = {10.1126/sciadv.1601832},
  urldate = {2021-08-10},
  langid = {english}
}

@article{Bao17NanoLett.,
  title = {Stacking-{{Dependent Electronic Structure}} of {{Trilayer Graphene Resolved}} by {{Nanospot Angle-Resolved Photoemission Spectroscopy}}},
  author = {Bao, Changhua and Yao, Wei and Wang, Eryin and Chen, Chaoyu and Avila, Jos{\'e} and Asensio, Maria C. and Zhou, Shuyun},
  year = {2017},
  month = mar,
  journal = {Nano Letters},
  volume = {17},
  number = {3},
  pages = {1564--1568},
  issn = {1530-6984, 1530-6992},
  doi = {10.1021/acs.nanolett.6b04698},
  urldate = {2020-08-13},
  langid = {english}
}

@article{Nguyen19Nature,
  title = {Visualizing Electrostatic Gating Effects in Two-Dimensional Heterostructures},
  author = {Nguyen, Paul V. and Teutsch, Natalie C. and Wilson, Nathan P. and Kahn, Joshua and Xia, Xue and Graham, Abigail J. and Kandyba, Viktor and Giampietri, Alessio and Barinov, Alexei and Constantinescu, Gabriel C. and Yeung, Nelson and Hine, Nicholas D. M. and Xu, Xiaodong and Cobden, David H. and Wilson, Neil R.},
  year = {2019},
  month = aug,
  journal = {Nature},
  volume = {572},
  number = {7768},
  pages = {220--223},
  publisher = {Nature Publishing Group},
  issn = {1476-4687},
  doi = {10.1038/s41586-019-1402-1},
  urldate = {2020-07-19},
  copyright = {2019 The Author(s), under exclusive licence to Springer Nature Limited},
  langid = {english}
}

@article{Bostwick10Science,
  title = {Observation of {{Plasmarons}} in {{Quasi-Freestanding Doped Graphene}}},
  author = {Bostwick, Aaron and Speck, Florian and Seyller, Thomas and Horn, Karsten and Polini, Marco and Asgari, Reza and MacDonald, Allan H. and Rotenberg, Eli},
  year = {2010},
  month = may,
  journal = {Science},
  volume = {328},
  number = {5981},
  pages = {999--1002},
  publisher = {American Association for the Advancement of Science},
  doi = {10.1126/science.1186489},
  urldate = {2024-12-04}
}

@article{Medjanik17NatureMater,
  title = {Direct {{3D}} Mapping of the {{Fermi}} Surface and {{Fermi}} Velocity},
  author = {Medjanik, K. and Fedchenko, O. and Chernov, S. and Kutnyakhov, D. and Ellguth, M. and Oelsner, A. and Sch{\"o}nhense, B. and Peixoto, T. R. F. and Lutz, P. and Min, C.-H. and Reinert, F. and D{\"a}ster, S. and Acremann, Y. and Viefhaus, J. and Wurth, W. and Elmers, H. J. and Sch{\"o}nhense, G.},
  year = {2017},
  month = jun,
  journal = {Nature Materials},
  volume = {16},
  number = {6},
  pages = {615--621},
  publisher = {Nature Publishing Group},
  issn = {1476-4660},
  doi = {10.1038/nmat4875},
  urldate = {2024-07-03},
  copyright = {2017 Springer Nature Limited},
  langid = {english}
}

@article{Kromker08Rev.Sci.Instrum.,
  title = {Development of a Momentum Microscope for Time Resolved Band Structure Imaging},
  author = {Kr{\"o}mker, B. and Escher, M. and Funnemann, D. and Hartung, D. and Engelhard, H. and Kirschner, J.},
  year = {2008},
  month = may,
  journal = {Review of Scientific Instruments},
  volume = {79},
  number = {5},
  pages = {053702},
  issn = {0034-6748, 1089-7623},
  doi = {10.1063/1.2918133},
  urldate = {2024-07-04},
  langid = {english}
}

@article{Madeo20Science,
  title = {Directly Visualizing the Momentum-Forbidden Dark Excitons and Their Dynamics in Atomically Thin Semiconductors},
  author = {Mad{\'e}o, Julien and Man, Michael K. L. and Sahoo, Chakradhar and Campbell, Marshall and Pareek, Vivek and Wong, E. Laine and {Al-Mahboob}, Abdullah and Chan, Nicholas S. and Karmakar, Arka and Mariserla, Bala Murali Krishna and Li, Xiaoqin and Heinz, Tony F. and Cao, Ting and Dani, Keshav M.},
  year = {2020},
  month = dec,
  journal = {Science},
  volume = {370},
  number = {6521},
  pages = {1199--1204},
  publisher = {American Association for the Advancement of Science},
  issn = {0036-8075, 1095-9203},
  doi = {10.1126/science.aba1029},
  urldate = {2021-03-03},
  chapter = {Report},
  copyright = {Copyright {\copyright} 2020 The Authors, some rights reserved; exclusive licensee American Association for the Advancement of Science. No claim to original U.S. Government Works. https://www.sciencemag.org/about/science-licenses-journal-article-reuseThis is an article distributed under the terms of the Science Journals Default License.},
  langid = {english},
  pmid = {33273099}
}

@article{Wallauer21NanoLett.,
  title = {Momentum-{{Resolved Observation}} of {{Exciton Formation Dynamics}} in {{Monolayer WS}}{\textsubscript{2}}},
  author = {Wallauer, Robert and {Perea-Causin}, Raul and M{\"u}nster, Lasse and Zajusch, Sarah and Brem, Samuel and G{\"u}dde, Jens and Tanimura, Katsumi and Lin, Kai-Qiang and Huber, Rupert and Malic, Ermin and H{\"o}fer, Ulrich},
  year = {2021},
  month = jul,
  journal = {Nano Letters},
  volume = {21},
  number = {13},
  pages = {5867--5873},
  issn = {1530-6984, 1530-6992},
  doi = {10.1021/acs.nanolett.1c01839},
  urldate = {2021-09-16},
  langid = {english}
}

@article{Schmitt22Nature,
  title = {Formation of Moir{\'e} Interlayer Excitons in Space and Time},
  author = {Schmitt, David and Bange, Jan Philipp and Bennecke, Wiebke and AlMutairi, AbdulAziz and Meneghini, Giuseppe and Watanabe, Kenji and Taniguchi, Takashi and Steil, Daniel and Luke, D. Russell and Weitz, R. Thomas and Steil, Sabine and Jansen, G. S. Matthijs and Brem, Samuel and Malic, Ermin and Hofmann, Stephan and Reutzel, Marcel and Mathias, Stefan},
  year = {2022},
  month = aug,
  journal = {Nature},
  volume = {608},
  number = {7923},
  pages = {499--503},
  publisher = {Nature Publishing Group},
  issn = {1476-4687},
  doi = {10.1038/s41586-022-04977-7},
  urldate = {2022-08-22},
  copyright = {2022 The Author(s), under exclusive licence to Springer Nature Limited},
  langid = {english}
}

@article{delaTorre21Rev.Mod.Phys.,
  title = {Colloquium: {{Nonthermal}} Pathways to Ultrafast Control in Quantum Materials},
  shorttitle = {Colloquium},
  author = {{de la Torre}, Alberto and Kennes, Dante M. and Claassen, Martin and Gerber, Simon and McIver, James W. and Sentef, Michael A.},
  year = {2021},
  month = oct,
  journal = {Reviews of Modern Physics},
  volume = {93},
  number = {4},
  pages = {041002},
  publisher = {American Physical Society},
  doi = {10.1103/RevModPhys.93.041002},
  urldate = {2025-05-22}
}

@article{Mak22Nat.Nanotechnol.,
  title = {Semiconductor Moir{\'e} Materials},
  author = {Mak, Kin Fai and Shan, Jie},
  year = {2022},
  month = jul,
  journal = {Nature Nanotechnology},
  volume = {17},
  number = {7},
  pages = {686--695},
  publisher = {Nature Publishing Group},
  issn = {1748-3395},
  doi = {10.1038/s41565-022-01165-6},
  urldate = {2025-02-11},
  copyright = {2022 Springer Nature Limited},
  langid = {english}
}

@article{Barrett12Rev.Sci.Instrum.,
  title = {Dark Field Photoelectron Emission Microscopy of Micron Scale Few Layer Graphene},
  author = {Barrett, N. and Conrad, E. and Winkler, K. and Kr{\"o}mker, B.},
  year = {2012},
  month = aug,
  journal = {Review of Scientific Instruments},
  volume = {83},
  number = {8},
  pages = {083706},
  issn = {0034-6748, 1089-7623},
  doi = {10.1063/1.4746279},
  urldate = {2022-04-01},
  langid = {english}
}

@article{Karni23Adv.Mater.,
  title = {Through the {{Lens}} of a {{Momentum Microscope}}: {{Viewing Light-Induced Quantum Phenomena}} in {{2D Materials}}},
  shorttitle = {Through the {{Lens}} of a {{Momentum Microscope}}},
  author = {Karni, Ouri and Esin, Iliya and Dani, Keshav M.},
  year = {2023},
  journal = {Advanced Materials},
  volume = {35},
  number = {27},
  pages = {2204120},
  issn = {1521-4095},
  doi = {10.1002/adma.202204120},
  urldate = {2024-07-10},
  copyright = {{\copyright} 2022 The Authors. Advanced Materials published by Wiley-VCH GmbH},
  langid = {english}
}

@article{Bange232DMater.,
  title = {Ultrafast Dynamics of Bright and Dark Excitons in Monolayer {{WSe}}{\textsubscript{2}} and Heterobilayer {{WSe}}{\textsubscript{2}}/{{MoS}}{\textsubscript{2}}},
  author = {Bange, Jan Philipp and Werner, Paul and Schmitt, David and Bennecke, Wiebke and Meneghini, Giuseppe and AlMutairi, AbdulAziz and Merboldt, Marco and Watanabe, Kenji and Taniguchi, Takashi and Steil, Sabine and Steil, Daniel and Weitz, R Thomas and Hofmann, Stephan and Jansen, G S Matthijs and Brem, Samuel and Malic, Ermin and Reutzel, Marcel and Mathias, Stefan},
  year = {2023},
  month = jul,
  journal = {2D Materials},
  volume = {10},
  number = {3},
  pages = {035039},
  issn = {2053-1583},
  doi = {10.1088/2053-1583/ace067},
  urldate = {2023-07-03},
  langid = {english}
}

@article{Bange24Sci.Adv.,
  title = {Probing Electron-Hole {{Coulomb}} Correlations in the Exciton Landscape of a Twisted Semiconductor Heterostructure},
  author = {Bange, Jan Philipp and Schmitt, David and Bennecke, Wiebke and Meneghini, Giuseppe and AlMutairi, AbdulAziz and Watanabe, Kenji and Taniguchi, Takashi and Steil, Daniel and Steil, Sabine and Weitz, R. Thomas and Jansen, G. S. Matthijs and Hofmann, Stephan and Brem, Samuel and Malic, Ermin and Reutzel, Marcel and Mathias, Stefan},
  year = {2024},
  month = feb,
  journal = {Science Advances},
  volume = {10},
  number = {6},
  pages = {eadi1323},
  publisher = {American Association for the Advancement of Science},
  doi = {10.1126/sciadv.adi1323},
  urldate = {2024-02-14}
}

@article{Dong21Nat.Sci.,
  title = {Direct Measurement of Key Exciton Properties: {{Energy}}, Dynamics, and Spatial Distribution of the Wave Function},
  shorttitle = {Direct Measurement of Key Exciton Properties},
  author = {Dong, Shuo and Puppin, Michele and Pincelli, Tommaso and Beaulieu, Samuel and Christiansen, Dominik and H{\"u}bener, Hannes and Nicholson, Christopher W. and Xian, Rui Patrick and Dendzik, Maciej and Deng, Yunpei and Windsor, Yoav William and Selig, Malte and Malic, Ermin and Rubio, Angel and Knorr, Andreas and Wolf, Martin and Rettig, Laurenz and Ernstorfer, Ralph},
  year = {2021},
  journal = {Natural Sciences},
  volume = {1},
  number = {1},
  pages = {e10010},
  issn = {2698-6248},
  doi = {10.1002/ntls.10010},
  urldate = {2021-12-15},
  langid = {english}
}

@article{Kunin23Phys.Rev.Lett.,
  title = {Momentum-{{Resolved Exciton Coupling}} and {{Valley Polarization Dynamics}} in {{Monolayer WS}}{\textsubscript{2}}},
  author = {Kunin, Alice and Chernov, Sergey and Bakalis, Jin and Li, Ziling and Cheng, Shuyu and Withers, Zachary H. and White, Michael G. and Sch{\"o}nhense, Gerd and Du, Xu and Kawakami, Roland K. and Allison, Thomas K.},
  year = {2023},
  month = jan,
  journal = {Physical Review Letters},
  volume = {130},
  number = {4},
  pages = {046202},
  publisher = {American Physical Society},
  doi = {10.1103/PhysRevLett.130.046202},
  urldate = {2023-01-30}
}

@article{Bennecke24NatCommun,
  title = {Disentangling the Multiorbital Contributions of Excitons by Photoemission Exciton Tomography},
  author = {Bennecke, Wiebke and Windischbacher, Andreas and Schmitt, David and Bange, Jan Philipp and Hemm, Ralf and Kern, Christian S. and D'Avino, Gabriele and Blase, Xavier and Steil, Daniel and Steil, Sabine and Aeschlimann, Martin and Stadtm{\"u}ller, Benjamin and Reutzel, Marcel and Puschnig, Peter and Jansen, G. S. Matthijs and Mathias, Stefan},
  year = {2024},
  month = feb,
  journal = {Nature Communications},
  volume = {15},
  number = {1},
  pages = {1804},
  publisher = {Nature Publishing Group},
  issn = {2041-1723},
  doi = {10.1038/s41467-024-45973-x},
  urldate = {2024-07-08},
  copyright = {2024 The Author(s)},
  langid = {english}
}

@article{Duvel22NanoLett.,
  title = {Far-from-{{Equilibrium Electron}}--{{Phonon Interactions}} in {{Optically Excited Graphene}}},
  author = {D{\"u}vel, Marten and Merboldt, Marco and Bange, Jan Philipp and Strauch, Hannah and Stellbrink, Michael and Pierz, Klaus and Schumacher, Hans Werner and Momeni, Davood and Steil, Daniel and Jansen, G. S. Matthijs and Steil, Sabine and Novko, Dino and Mathias, Stefan and Reutzel, Marcel},
  year = {2022},
  month = jun,
  journal = {Nano Letters},
  volume = {22},
  number = {12},
  pages = {4897--4904},
  publisher = {American Chemical Society},
  issn = {1530-6984},
  doi = {10.1021/acs.nanolett.2c01325},
  urldate = {2022-10-20}
}

@article{Mueller18npj2DMaterAppl,
  title = {Exciton Physics and Device Application of Two-Dimensional Transition Metal Dichalcogenide Semiconductors},
  author = {Mueller, Thomas and Malic, Ermin},
  year = {2018},
  month = sep,
  journal = {npj 2D Materials and Applications},
  volume = {2},
  number = {1},
  pages = {29},
  publisher = {Nature Publishing Group},
  issn = {2397-7132},
  doi = {10.1038/s41699-018-0074-2},
  urldate = {2024-05-07},
  copyright = {2018 The Author(s)},
  langid = {english}
}

@article{Ulstrup19NatCommun,
  title = {Nanoscale Mapping of Quasiparticle Band Alignment},
  author = {Ulstrup, S{\o}ren and Giusca, Cristina E. and Miwa, Jill A. and Sanders, Charlotte E. and Browning, Alex and Dudin, Pavel and Cacho, Cephise and Kazakova, Olga and Gaskill, D. Kurt and {Myers-Ward}, Rachael L. and Zhang, Tianyi and Terrones, Mauricio and Hofmann, Philip},
  year = {2019},
  month = jul,
  journal = {Nature Communications},
  volume = {10},
  number = {1},
  pages = {3283},
  publisher = {Nature Publishing Group},
  issn = {2041-1723},
  doi = {10.1038/s41467-019-11253-2},
  urldate = {2024-02-14},
  copyright = {2019 The Author(s)},
  langid = {english}
}

@article{Schonhense18NewJ.Phys.,
  title = {Multidimensional Photoemission Spectroscopy---the Space-Charge Limit},
  author = {Sch{\"o}nhense, B and Medjanik, K and Fedchenko, O and Chernov, S and Ellguth, M and Vasilyev, D and Oelsner, A and Viefhaus, J and Kutnyakhov, D and Wurth, W and Elmers, H J and Sch{\"o}nhense, G},
  year = {2018},
  month = mar,
  journal = {New Journal of Physics},
  volume = {20},
  number = {3},
  pages = {033004},
  publisher = {IOP Publishing},
  issn = {1367-2630},
  doi = {10.1088/1367-2630/aaa262},
  urldate = {2025-08-12},
  langid = {english}
}

@article{Keunecke20Rev.Sci.Instrum.,
  title = {Time-Resolved Momentum Microscopy with a 1 {{MHz}} High-Harmonic Extreme Ultraviolet Beamline},
  author = {Keunecke, Marius and M{\"o}ller, Christina and Schmitt, David and Nolte, Hendrik and Jansen, G. S. Matthijs and Reutzel, Marcel and Gutberlet, Marie and Halasi, Gyula and Steil, Daniel and Steil, Sabine and Mathias, Stefan},
  year = {2020},
  month = jun,
  journal = {Review of Scientific Instruments},
  volume = {91},
  number = {6},
  pages = {063905},
  publisher = {American Institute of Physics},
  issn = {0034-6748},
  doi = {10.1063/5.0006531},
  urldate = {2021-03-04}
}

@article{Puppin19Rev.Sci.Instrum.,
  title = {Time- and Angle-Resolved Photoemission Spectroscopy of Solids in the Extreme Ultraviolet at 500 {{kHz}} Repetition Rate},
  author = {Puppin, M. and Deng, Y. and Nicholson, C. W. and Feldl, J. and Schr{\"o}ter, N. B. M. and Vita, H. and Kirchmann, P. S. and Monney, C. and Rettig, L. and Wolf, M. and Ernstorfer, R.},
  year = {2019},
  month = feb,
  journal = {Review of Scientific Instruments},
  volume = {90},
  number = {2},
  pages = {023104},
  issn = {0034-6748, 1089-7623},
  doi = {10.1063/1.5081938},
  urldate = {2019-06-27},
  langid = {english}
}

@article{Allison25APLPhotonics,
  title = {Cavity-Enhanced High-Order Harmonic Generation for High-Performance Time-Resolved Photoemission Experiments},
  author = {Allison, Thomas K. and Kunin, Alice and Sch{\"o}nhense, Gerd},
  year = {2025},
  month = jan,
  journal = {APL Photonics},
  volume = {10},
  number = {1},
  pages = {010906},
  issn = {2378-0967},
  doi = {10.1063/5.0244045},
  urldate = {2025-08-12}
}

@article{Gerber17Science,
  title = {Femtosecond Electron-Phonon Lock-in by Photoemission and x-Ray Free-Electron Laser},
  author = {Gerber, S. and Yang, S.-L. and Zhu, D. and Soifer, H. and Sobota, J. A. and Rebec, S. and Lee, J. J. and Jia, T. and Moritz, B. and Jia, C. and Gauthier, A. and Li, Y. and Leuenberger, D. and Zhang, Y. and Chaix, L. and Li, W. and Jang, H. and Lee, J.-S. and Yi, M. and Dakovski, G. L. and Song, S. and Glownia, J. M. and Nelson, S. and Kim, K. W. and Chuang, Y.-D. and Hussain, Z. and Moore, R. G. and Devereaux, T. P. and Lee, W.-S. and Kirchmann, P. S. and Shen, Z.-X.},
  year = {2017},
  month = jul,
  journal = {Science},
  volume = {357},
  number = {6346},
  pages = {71--75},
  publisher = {American Association for the Advancement of Science},
  doi = {10.1126/science.aak9946},
  urldate = {2025-08-12}
}

@article{Stojchevska14Science,
  title = {Ultrafast {{Switching}} to a {{Stable Hidden Quantum State}} in an {{Electronic Crystal}}},
  author = {Stojchevska, L. and Vaskivskyi, I. and Mertelj, T. and Kusar, P. and Svetin, D. and Brazovskii, S. and Mihailovic, D.},
  year = {2014},
  month = apr,
  journal = {Science},
  volume = {344},
  number = {6180},
  pages = {177--180},
  publisher = {American Association for the Advancement of Science},
  doi = {10.1126/science.1241591},
  urldate = {2025-08-12}
}

@article{Arsenault24Phys.Rev.Lett.,
  title = {Two-{{Dimensional Moir{\'e} Polaronic Electron Crystals}}},
  author = {Arsenault, Eric A. and Li, Yiliu and Yang, Birui and Wang, Xi and Park, Heonjoon and Mosconi, Edoardo and Ronca, Enrico and Taniguchi, Takashi and Watanabe, Kenji and Gamelin, Daniel and Millis, Andrew and Dean, Cory R. and De Angelis, Filippo and Xu, Xiaodong and Zhu, X. Y.},
  year = {2024},
  month = mar,
  journal = {Physical Review Letters},
  volume = {132},
  number = {12},
  pages = {126501},
  issn = {0031-9007, 1079-7114},
  doi = {10.1103/PhysRevLett.132.126501},
  urldate = {2024-06-27},
  langid = {english}
}

@article{Schonhense15JournalofElectronSpectroscopyandRelatedPhenomena,
  title = {Space-, Time- and Spin-Resolved Photoemission},
  author = {Sch{\"o}nhense, Gerd and Medjanik, Katerina and Elmers, Hans-Joachim},
  year = {2015},
  month = apr,
  journal = {Journal of Electron Spectroscopy and Related Phenomena},
  series = {Special {{Anniversary Issue}}: {{Volume}} 200},
  volume = {200},
  pages = {94--118},
  issn = {0368-2048},
  doi = {10.1016/j.elspec.2015.05.016},
  urldate = {2025-08-13}
}

@article{Schonhense17Ultramicroscopy,
  title = {Spin-Filtered Time-of-Flight {\textbf{k}}-Space Microscopy of {{Ir}} -- {{Towards}} the ``Complete'' Photoemission Experiment},
  author = {Sch{\"o}nhense, G. and Medjanik, K. and Chernov, S. and Kutnyakhov, D. and Fedchenko, O. and Ellguth, M. and Vasilyev, D. and {Zaporozhchenko-Zymakov{\'a}}, A. and Panzer, D. and Oelsner, A. and Tusche, C. and Sch{\"o}nhense, B. and Braun, J. and Min{\'a}r, J. and Ebert, H. and Viefhaus, J. and Wurth, W. and Elmers, H. J.},
  year = {2017},
  month = dec,
  journal = {Ultramicroscopy},
  series = {{{LEEM}}/{{PEEM-10}}},
  volume = {183},
  pages = {19--29},
  issn = {0304-3991},
  doi = {10.1016/j.ultramic.2017.06.025},
  urldate = {2025-08-13}
}

@article{Reutzel24Adv.Phys.X,
  title = {Probing Excitons with Time-Resolved Momentum Microscopy},
  author = {Reutzel, Marcel and Jansen, G. S. Matthijs and Mathias, Stefan},
  year = {2024},
  month = dec,
  journal = {Advances in Physics: X},
  volume = {9},
  number = {1},
  pages = {2378722},
  publisher = {Taylor \& Francis},
  issn = {null},
  doi = {10.1080/23746149.2024.2378722},
  urldate = {2024-10-07}
}

@article{Keimer17NaturePhys,
  title = {The Physics of Quantum Materials},
  author = {Keimer, B. and Moore, J. E.},
  year = {2017},
  month = nov,
  journal = {Nature Physics},
  volume = {13},
  number = {11},
  pages = {1045--1055},
  publisher = {Nature Publishing Group},
  issn = {1745-2481},
  doi = {10.1038/nphys4302},
  urldate = {2025-07-24},
  copyright = {2017 Springer Nature Limited},
  langid = {english}
}

@article{Kennes21Nat.Phys.,
  title = {Moir{\'e} Heterostructures as a Condensed-Matter Quantum Simulator},
  author = {Kennes, Dante M. and Claassen, Martin and Xian, Lede and Georges, Antoine and Millis, Andrew J. and Hone, James and Dean, Cory R. and Basov, D. N. and Pasupathy, Abhay N. and Rubio, Angel},
  year = {2021},
  month = feb,
  journal = {Nature Physics},
  volume = {17},
  number = {2},
  pages = {155--163},
  publisher = {Nature Publishing Group},
  issn = {1745-2481},
  doi = {10.1038/s41567-020-01154-3},
  urldate = {2024-05-07},
  copyright = {2021 Springer Nature Limited},
  langid = {english}
}

@article{Rossnagel11J.Phys.:Condens.Matter,
  title = {On the Origin of Charge-Density Waves in Select Layered Transition-Metal Dichalcogenides},
  author = {Rossnagel, K},
  year = {2011},
  month = jun,
  journal = {Journal of Physics: Condensed Matter},
  volume = {23},
  number = {21},
  pages = {213001},
  issn = {0953-8984, 1361-648X},
  doi = {10.1088/0953-8984/23/21/213001},
  urldate = {2025-02-12},
  langid = {english}
}

@article{Dabrowski20Chem.Rev.,
  title = {Ultrafast {{Photoemission Electron Microscopy}}: {{Imaging Plasmons}} in {{Space}} and {{Time}}},
  shorttitle = {Ultrafast {{Photoemission Electron Microscopy}}},
  author = {D{\c a}browski, Maciej and Dai, Yanan and Petek, Hrvoje},
  year = 2020,
  month = jul,
  journal = {Chemical Reviews},
  volume = {120},
  number = {13},
  pages = {6247--6287},
  publisher = {American Chemical Society},
  issn = {0009-2665},
  doi = {10.1021/acs.chemrev.0c00146},
  urldate = {2026-01-14}
}

@book{Petek12,
  title = {Dynamics at Solid State Surfaces and Interfaces: {{Volume}} 2: {{Fundamentals}}},
  shorttitle = {Dynamics at Solid State Surfaces and Interfaces},
  editor = {Petek, Hrvoje and Wolf, Martin and Bovensiepen, Uwe},
  year = {2012},
  publisher = {John Wiley},
  address = {Chichester},
  isbn = {978-3-527-40924-2}
}

@article{Raja19Nat.Nanotechnol.,
  title = {Dielectric Disorder in Two-Dimensional Materials},
  author = {Raja, Archana and Waldecker, Lutz and Zipfel, Jonas and Cho, Yeongsu and Brem, Samuel and Ziegler, Jonas D. and Kulig, Marvin and Taniguchi, Takashi and Watanabe, Kenji and Malic, Ermin and Heinz, Tony F. and Berkelbach, Timothy C. and Chernikov, Alexey},
  year = {2019},
  month = sep,
  journal = {Nature Nanotechnology},
  volume = {14},
  number = {9},
  pages = {832--837},
  publisher = {Nature Publishing Group},
  issn = {1748-3395},
  doi = {10.1038/s41565-019-0520-0},
  urldate = {2022-08-02},
  copyright = {2019 The Author(s), under exclusive licence to Springer Nature Limited},
  langid = {english}
}

@article{Fedchenko22J.Phys.:Condens.Matter,
  title = {Direct Observation of Antiferromagnetic Parity Violation in the Electronic Structure of {{Mn}}{\textsubscript{2}}{{Au}}},
  author = {Fedchenko, O. and {\v S}mejkal, L. and Kallmayer, M. and Lytvynenko, Ya and Medjanik, K. and Babenkov, S. and Vasilyev, D. and Kl{\"a}ui, M. and Demsar, J. and Sch{\"o}nhense, G. and Jourdan, M. and Sinova, J. and Elmers, H. J.},
  year = {2022},
  month = aug,
  journal = {Journal of Physics: Condensed Matter},
  volume = {34},
  number = {42},
  pages = {425501},
  publisher = {IOP Publishing},
  issn = {0953-8984},
  doi = {10.1088/1361-648X/ac87e6},
  urldate = {2024-11-14},
  langid = {english}
}

@article{Hellmann12NatCommun,
  title = {Time-Domain Classification of Charge-Density-Wave Insulators},
  author = {Hellmann, S. and Rohwer, T. and Kall{\"a}ne, M. and Hanff, K. and Sohrt, C. and Stange, A. and Carr, A. and Murnane, M. M. and Kapteyn, H. C. and Kipp, L. and Bauer, M. and Rossnagel, K.},
  year = {2012},
  month = sep,
  journal = {Nature Communications},
  volume = {3},
  number = {1},
  pages = {1069},
  issn = {2041-1723},
  doi = {10.1038/ncomms2078},
  urldate = {2019-12-06},
  copyright = {2012 Nature Publishing Group, a division of Macmillan Publishers Limited. All Rights Reserved.},
  langid = {english}
}

@article{Xu25npjQuantumMater.,
  title = {Time-Domain Study of Coupled Collective Excitations in Quantum Materials},
  author = {Xu, Chenhang and Zong, Alfred},
  year = {2025},
  month = feb,
  journal = {npj Quantum Materials},
  volume = {10},
  number = {1},
  pages = {21},
  publisher = {Nature Publishing Group},
  issn = {2397-4648},
  doi = {10.1038/s41535-025-00726-x},
  urldate = {2025-08-19},
  copyright = {2025 The Author(s)},
  langid = {english}
}

@article{Feist17Ultramicroscopy,
  title = {Ultrafast Transmission Electron Microscopy Using a Laser-Driven Field Emitter: {{Femtosecond}} Resolution with a High Coherence Electron Beam},
  shorttitle = {Ultrafast Transmission Electron Microscopy Using a Laser-Driven Field Emitter},
  author = {Feist, Armin and Bach, Nora and {Rubiano da Silva}, Nara and Danz, Thomas and M{\"o}ller, Marcel and Priebe, Katharina E. and Domr{\"o}se, Till and Gatzmann, J. Gregor and Rost, Stefan and Schauss, Jakob and Strauch, Stefanie and Bormann, Reiner and Sivis, Murat and Sch{\"a}fer, Sascha and Ropers, Claus},
  year = {2017},
  month = may,
  journal = {Ultramicroscopy},
  series = {70th {{Birthday}} of {{Robert Sinclair}} and 65th {{Birthday}} of {{Nestor J}}. {{Zaluzec PICO}} 2017 -- {{Fourth Conference}} on {{Frontiers}} of {{Aberration Corrected Electron Microscopy}}},
  volume = {176},
  pages = {63--73},
  issn = {0304-3991},
  doi = {10.1016/j.ultramic.2016.12.005},
  urldate = {2025-08-19}
}

@article{Cocker16Nature,
  title = {Tracking the Ultrafast Motion of a Single Molecule by Femtosecond Orbital Imaging},
  author = {Cocker, Tyler L. and Peller, Dominik and Yu, Ping and Repp, Jascha and Huber, Rupert},
  year = {2016},
  month = nov,
  journal = {Nature},
  volume = {539},
  number = {7628},
  pages = {263--267},
  publisher = {Nature Publishing Group},
  issn = {1476-4687},
  doi = {10.1038/nature19816},
  urldate = {2025-08-19},
  copyright = {2016 Macmillan Publishers Limited, part of Springer Nature. All rights reserved.},
  langid = {english}
}

@article{Weinelt04Phys.Rev.Lett.,
  title = {Dynamics of {{Exciton Formation}} at the {{Si}}(100) c(4x2) {{Surface}}},
  author = {Weinelt, Martin and Kutschera, Michael and Fauster, Thomas and Rohlfing, Michael},
  year = {2004},
  month = mar,
  journal = {Physical Review Letters},
  volume = {92},
  number = {12},
  pages = {126801},
  publisher = {American Physical Society},
  doi = {10.1103/PhysRevLett.92.126801},
  urldate = {2023-01-11}
}

@article{Bauer94Rep.Prog.Phys.,
  title = {Low Energy Electron Microscopy},
  author = {Bauer, E.},
  year = {1994},
  month = sep,
  journal = {Reports on Progress in Physics},
  volume = {57},
  number = {9},
  pages = {895},
  issn = {0034-4885},
  doi = {10.1088/0034-4885/57/9/002},
  urldate = {2025-08-26},
  langid = {english}
}

@article{Regan22NatRevMater,
  title = {Emerging Exciton Physics in Transition Metal Dichalcogenide Heterobilayers},
  author = {Regan, Emma C. and Wang, Danqing and Paik, Eunice Y. and Zeng, Yongxin and Zhang, Long and Zhu, Jihang and MacDonald, Allan H. and Deng, Hui and Wang, Feng},
  year = {2022},
  month = oct,
  journal = {Nature Reviews Materials},
  volume = {7},
  number = {10},
  pages = {778--795},
  publisher = {Nature Publishing Group},
  issn = {2058-8437},
  doi = {10.1038/s41578-022-00440-1},
  urldate = {2023-04-13},
  copyright = {2022 Springer Nature Limited},
  langid = {english}
}

\end{document}